\documentclass[altaffilletter,superscriptaddress,amsmath,amssymb,twocolumn]{revtex4-1}
\usepackage{graphicx}
\usepackage{natbib}

\begin{document}

\title{On-chip, self-detected THz dual-comb spectrometer}%

\author{Markus R\"osch}%
\email{mroesch@phys.ethz.ch}
\author{Giacomo Scalari}
\email{scalari@phys.ethz.ch}
\author{Gustavo Villares}
\author{Lorenzo Bosco}
\author{Mattias Beck}
\author{J\'er\^ome Faist}
\affiliation{ETH Zurich, Institute of Quantum Electronics, Auguste-Piccard-Hof 1, Zurich 8093, Switzerland}

\begin{abstract}
We present a directly generated on-chip dual-comb source at THz frequencies. The multi-heterodyne beating signal of two free-running THz quantum cascade laser frequency combs is measured electrically using one of the combs as a detector, fully exploiting the unique characteristics of quantum cascade active regions. Up to 30 modes can be detected corresponding to a spectral bandwidth of 630 GHz, being the available  bandwidth of the dual comb  configuration.  The multi-heterodyne signal is used to investigate the equidistance of the comb modes showing an accuracy of $10^{-12}$ at the carrier frequency of 2.5 THz. 
\end{abstract}

\maketitle 

Many substances have a characteristic molecular fingerprint at THz frequencies such as biomolecules, medicines, drugs, explosives, cancer tissue, DNA, proteins, or bacteria \cite{tonouchi2007,fischer2005,chen2005,fan2007,Burnett2009,leahy2009}, making this frequency range attractive for molecular spectroscopy applications. Currently most spectroscopy experiments at THz frequencies are performed using THz time-domain spectroscopy systems \cite{TDS_book_2005,tonouchi2007}. These systems are limited in power, having their strongest signal normally well below 2.5 THz. Another approach - especially for high-precision, and broadband THz spectroscopy - is dual-comb spectroscopy \cite{schiller2002spectrometry,keilmann2004time,yasui2006,bernhardt2010mid,villares2014,pavelyev2014,finneran2015,tammaro2015}. A dual-comb setup has no moving parts and in addition allows a very precise measurement of the frequency. Since in existing THz dual-comb setups the THz signal is not generated directly, they still present a wide footprint and low THz powers \cite{yasui2006,finneran2015,pavelyev2014,tammaro2015}. \\
In contrast, terahertz quantum cascade lasers (THz QCLs) are a very promising technology for applications at THz frequencies \cite{kohler2002,williams2007,tonouchi2007}.
THz QCLs have the advantage that they provide much more power, operate in continuous wave (CW), and extend towards higher frequencies \cite{williams2007,tonouchi2007}. A promising way to use quantum cascade technology for broadband, high-resolution spectroscopy in the THz is frequency comb operation \cite{Hugi2012,villares2014,Roesch2014,Burghoff2014,Wienold2014,Faist2015combs}. QCL combs can be realized by Four-wave mixing (FWM), due to the fast dynamics and the high non-linearity of the quantum cascade active region \cite{Faist2015combs}. Contrary to ultrafast mode-locked combs this will generate an almost constant output in the time-domain \cite{Faist2015combs,Khurgin2014}.\\ 
\begin{figure}[b]
  \centering
  \includegraphics[scale=1]{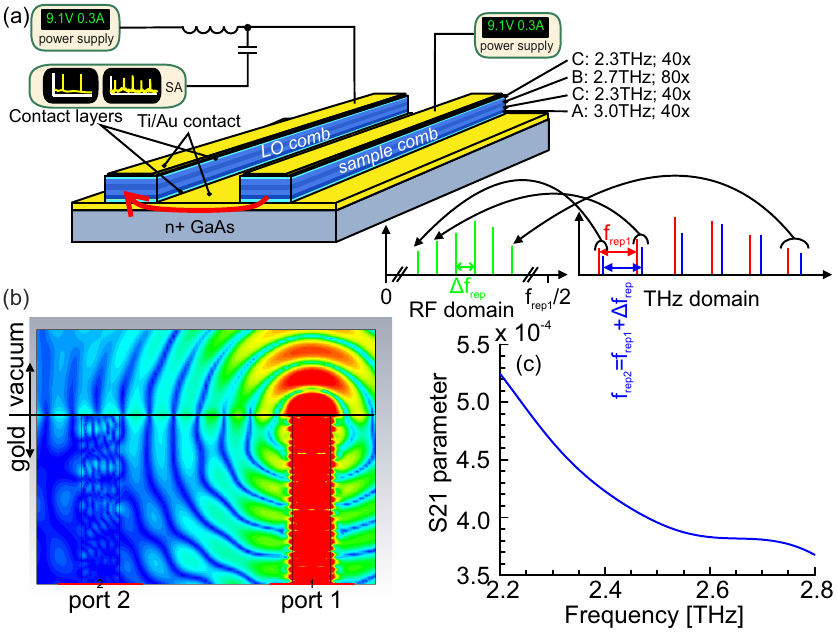}
  \caption{(a) Schematics of the dual-comb on chip. The lasers have 3 different active region designs as reported in reference \cite{Roesch2014}. The red arrow indicates the part of the light coupling from the sample comb to the local oscillator (LO) comb. The multi-heterodyne signal as well as the intermode beatnotes of both lasers can be extracted with a bias-tee on the current line of the LO comb. The graph on the right explains the relation between the optical comb spectra and the multi-heterodyne spectrum. (b) Finite element simulation of the dual-comb on chip. A broad frequency (2-3 THz) signal is launched from waveguide port (1) into the sample ridge. Part of the light couples into the LO ridge reaching port (2). The colorplot shows the absolute value of the electric field. (c) The S21 parameter is displayed indicating the fraction of E-field coupling from port 1 to port 2. All simulations were performed with \textsc{CST Microwave Studio$^{\circledR}$}.}
  \label{fig:simulations}
\end{figure}
\begin{figure*}[t]
  \centering
  \includegraphics[scale=1]{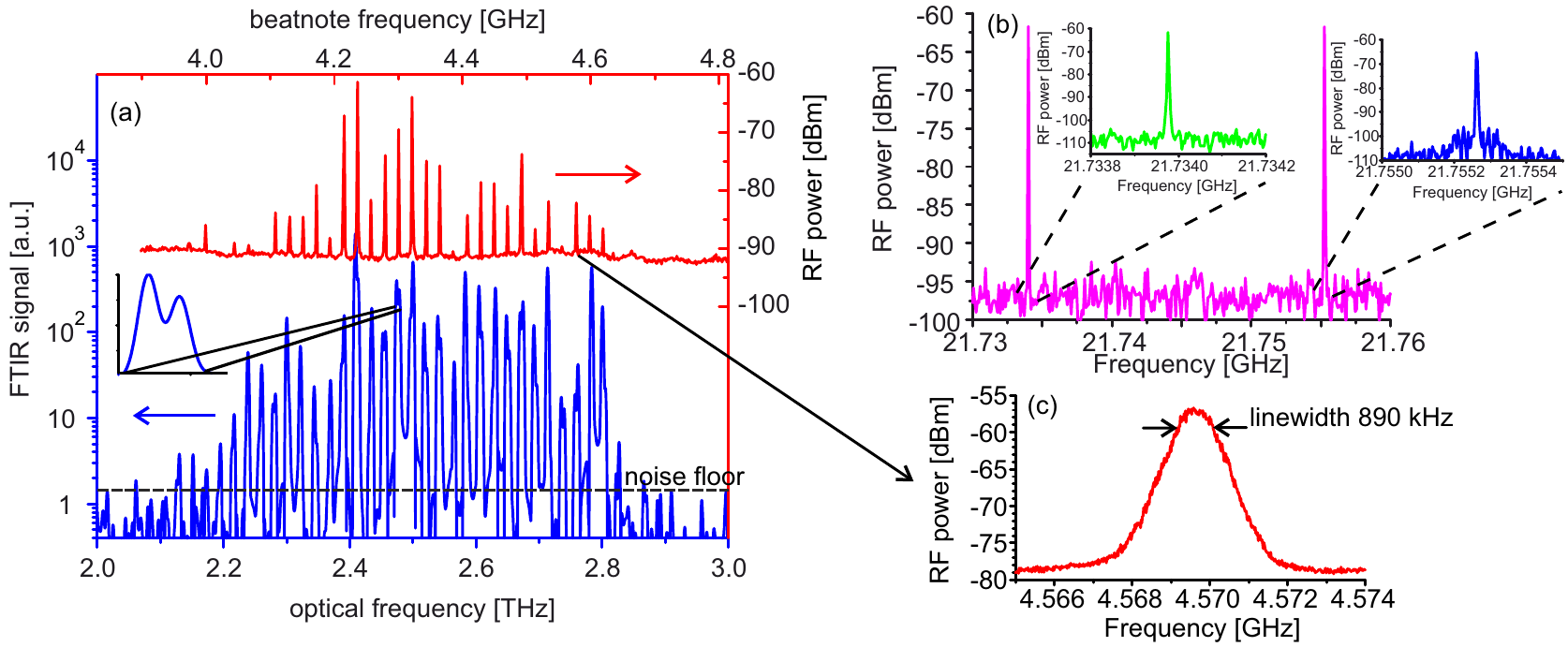}
  \caption{Dual-comb spectra: (a) optical spectrum (blue curve) of the dual-comb measured with a Bruker 80v FTIR. Laser 1 is driven at 318mA, laser 2 at 321mA. The temperature is set to 23 Kelvin. The inset shows that the modes consist of two peaks separated by 4.4 GHz close to the resolution limit of the FTIR (2.25GHz) corresponding to the two combs. In red is the corresponding multi-heterodyne spectrum extracted from the current bias of the LO laser (RBW=5 kHz, VBW=50 kHz, 40 sec sweep time (SWT)) (b) corresponding intermode beatnote signal (RBW=10 kHz, VBW= 100 kHz, SWT=120 ms). The insets show zooms of the individual beatnotes showing linewidths of below 1 kHz (RBW=1 kHz, VBW=10 kHz, SWT=40 ms) (c) Zoom of one line of the multi-heterodyne spectrum (laser 1: 307mA, laser 2: 323mA, Temperature: 23K, RBW=100 kHz, VBW=1 MHz, SWT=20 ms  averaged over 100 sweeps)}
  \label{fig:heterodyne}
\end{figure*}
Recent work using mid-IR QCL combs has shown that it is possible to realize a dual-comb experiment \cite{villares2014,villares2015dualcombonchip}. In such an experiment two frequency combs with slightly different mode-spacings are sent onto a fast detector. This detector will then detect the multi-heterodyne beating of the two combs (see figure \ref{fig:simulations}(a)). The main limitation for adapting a dual-comb setup to the THz region is the lack of sensitive fast detectors. We therefore developed a dual-comb setup which does not require a fast detector. In our approach, we are fully exploiting some of the unique features of QCLs since one of the QCL combs can be used as a detector itself. The two combs can be fabricated on a single chip reducing the footprint of the setup and getting rid of any additional optical components. Furthermore, the two combs will share the same thermal fluctuations and other technical noise increasing the stability of the multi-heterodyne signal.\\
We have recently reported a THz QCL featuring a regime where a single narrow beatnote was observed \cite{Roesch2014}. Due to theoretical models and other results on QCL based combs such a narrow beatnote is a strong indicator for frequency comb operation \cite{Hugi2012,Khurgin2014,Burghoff2014}, making this laser the ideal candidate for a THz dual-comb. Two lasers were processed by dry-etching techniques next to each other on the same chip and connected individually to two  current sources (see figure \ref{fig:simulations}(a) for details). The laser active region used is the one reported in reference \cite{Roesch2014}. The two lasers have the same width (50 $\mu$m), and are spaced by 450 $\mu$m. The cavities were defined by cleaving the ridges on the same chip to a length of 1.9 mm. \\
\begin{figure*}[tb]
  \centering
  \includegraphics[scale=1]{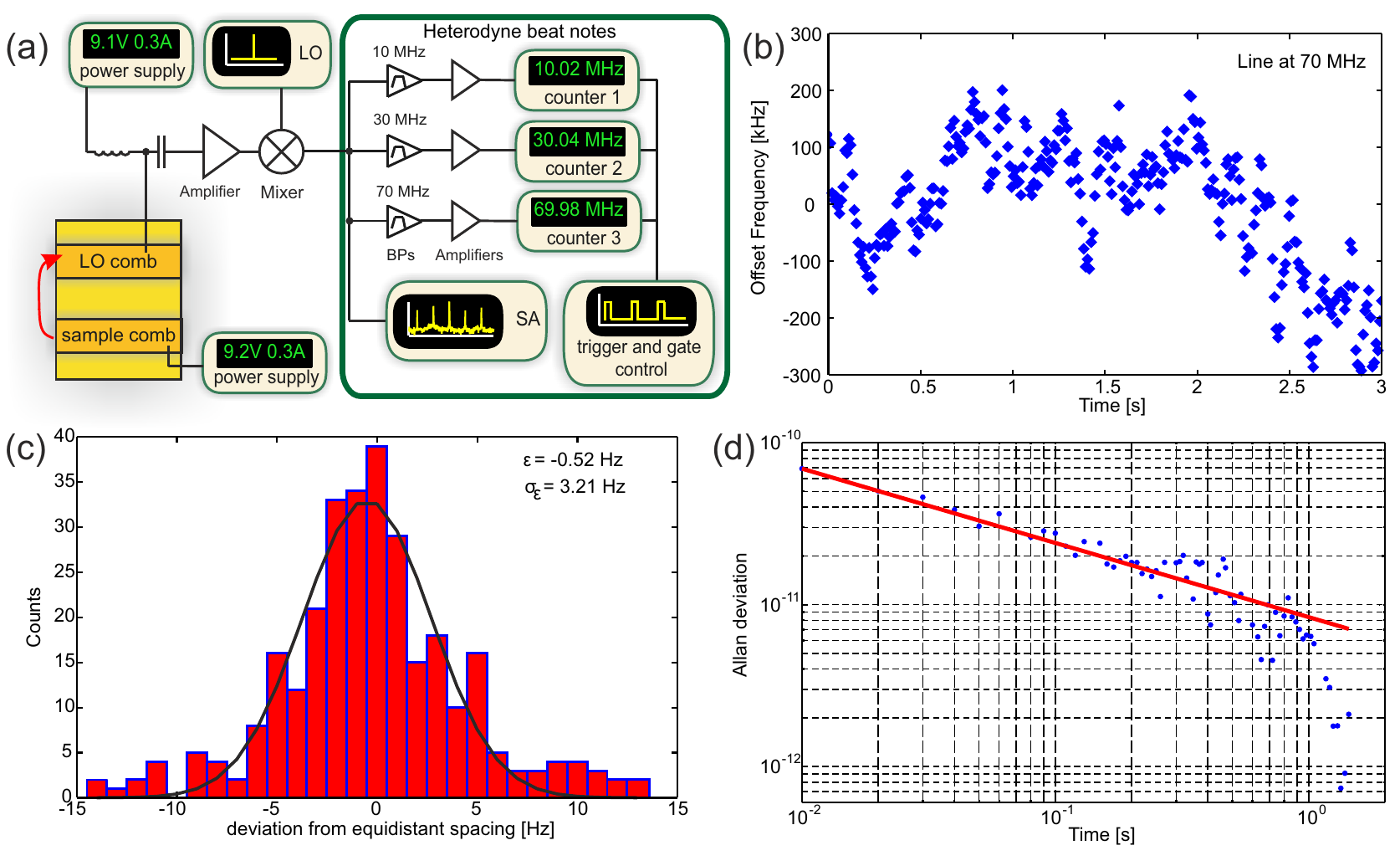}
  \caption{Equidistance measurement: (a) Schematics of the experimental setup (BP=bandpass filter, SA=spectrum analyzer, LO=local oscillator). The multi-heterodyne signal is extracted from the LO comb and down-mixed to below 200 MHz. Three lines at 10, 20, and 70 MHz are measured with three frequency counters (see text for details.) (b) Frequency fluctuations over time of the 70 MHz line from the down-mixed multi-heterodyne spectrum. The frequency was measured with a frequency counter (gate time=10ms). (c) Distribution histogram of the deviation from equidistant mode spacing $\epsilon$. The total measurement time was 3 seconds, gate time 10ms. The distribution has been fitted with a Gaussian function showing an average value of -0.52 Hz and a standard deviation of 3.21 Hz. (d) Allan deviation normalized to the measurement bandwidth (3*FSR=65.1 GHz), showing an inverse square root dependency.}
  \label{fig:HDBN}
\end{figure*}
Finite element simulations (\textsc{CST Microwave Studio}$^{\circledR}$) in figure \ref{fig:simulations}(b) show that part of the THz electric field exiting the laser cavity travels along the edge of the bottom metalization, and eventually couples into the neighboring laser cavity. In figure \ref{fig:simulations}(c) the S21 parameter is plotted. The S21 parameter gives the ratio of electric field launched from port (1) in the sample comb that couples into the local oscillator (LO) comb reaching port (2) (see also figure \ref{fig:simulations}(a),(b). The amount of power coupled to the LO laser is very small (of the order of 1-10 nW assuming an intra-cavity power of 5 mW and taking both facets into account). However, small amounts of light can perturb the laser bias \cite{green2008,dean2011,Wienold2014}. The ultrafast dynamics allows the THz QCL to act as an efficient  ultrafast detector, bypassing the need for an integrated Schottky diode as was done in ref.\cite{wanke2010}. THz QCLs are very efficient in detecting light as the electron dynamics is affected by photon-driven transport \cite{faist2013book}. The upper-state lifetime is very short and therefore very fast modulations can still be detected. So far this peculiarity has mainly been used to do imaging using the feedback reflected on some object in front of the laser \cite{dean2011,rakic2013}. In an other experiment the feedback from an Fourier-Transform Interferometer (FTIR) was used to probe the comb characteristics of a THz QCL \cite{Wienold2014}. In our case the radiation  of the neighbor  laser (sample laser) is coupled in the detector laser which acts also as LO for the multi-heterodyne signal (figure \ref{fig:simulations}(a)). The multi-heterodyne mixing of the two laser signals is then visible as a fast modulation on the laser bias of the LO laser. Extracting the fast oscillating part by a bias-tee one gets an electrical RF signal containing the multi-heterodyne beating of the two lasers. In addition, also the intermode beating signal of both lasers corresponding to their repetition frequencies is captured. \\
Both lasers show several regions with a single narrow beatnote \cite{footnote_BN}. Regimes can be found with both lasers featuring a single narrow ($<$ 1kHz) beatnote (figure \ref{fig:heterodyne}(b)). Both lasers are in a free running condition and only the temperature is stabilized at 23 Kelvin. Optical and multi-heterodyne spectra have been measured. Overlaying the optical and the multi-heterodyne spectra shows a remarkably good agreement (figure \ref{fig:heterodyne}(a)). A maximum-likelihood estimation has been implemented to match the multi-heterodyne and the optical spectra. The optical spectrum contains two sets of modes spaced by 4.4 GHz corresponding to the two combs (inset figure \ref{fig:heterodyne}(a)). The multi-heterodyne spectrum spans over 30 modes corresponding to 630 GHz bandwidth in the optical spectrum. There was no need to actively stabilize the combs and also no post-data treating was necessary to reveal the multi-heterodyne spectrum. The modes of the multi-heterodyne spectrum show, even without stabilization, jitter-limited linewidths of below 900 kHz (figure \ref{fig:heterodyne}(c)).\\
This experiment shows the power and simplicity of broadband, heterogeneous QCL combs. Compared to other THz dual-combs \cite{yasui2006,finneran2015}, the footprint of this QCL dual-comb is extremely small (a few $mm^2$), as no detector or optics is needed. Our experiment shows that already a small fraction of radiation can be detected by the  QCL  LO comb.\\
To further characterize the dual-comb, we present here an experiment which rigorously proves the equidistance of the mode spacing of our combs. In previous work on QCL combs, the so-called beatnote spectroscopy or shifted wave interference Fourier-transform spectroscopy were performed to proof the comb nature of the corresponding lasers \cite{Hugi2012,Burghoff2014}. A different approach is to use a dual-comb setup and perform a multi-heterodyne mixing on a fast detector \cite{schiller2002spectrometry,delhaye2007,villares2014}. The multi-heterodyne beating signal will consist of equally spaced modes corresponding to the difference in the modespacing of the two combs $\Delta f_{rep}=f_{rep_1}-f_{rep_2}$ (figure 1a). Each mode of the multi-heterodyne signal corresponds to a pair of modes, one from each comb. 
If one can proof that the multi-heterodyne modes are equally spaced, this immediately implies that also the two combs have perfectly equally spaced modes.\\
To proof the equidistance of the modes we use the same approach as reported in ref. \cite{delhaye2007,villares2014}, using frequency counting techniques available at RF frequencies. A few adaptations had to be made in order to make it suitable for our configuration. The full schematics of the experimental setup is displayed in figure \ref{fig:HDBN}(a). The multi-heterodyne signal (red curve in figure \ref{fig:heterodyne}(a)) is down-mixed to below 200 MHz with a RF mixer (Miteq DB0218HW2-R) and a local oscillator (Rodhe \& Schwarz SMF100A) to make it compatible with the used frequency counters. The combs have been adjusted to get a $\Delta f_{rep}$ of 20 MHz. The local oscillator was tuned to get the lowest frequency mode to lay at 10 MHz. The resulting signal is splitted into 4 parts. One is monitored on a spectrum analyzer (Rodhe \& Schwarz FSU50). The other three lines are filtered with three different narrow (2-3 MHz) bandpass filters (10,30,70 MHz, A-INFO: WBLB-T-BP-10-2-L,WBLB-T-BP-30-3-L,WBLB-T-BP-70-3-L). Each line is again amplified (Miteq AU-1519) before sending it to a frequency counter (Agilent 53220A). The three frequency counters share the same trigger, gate, and reference. One frequency counter is used as master providing the time reference and sharing it with the two other counters. Trigger and gate signal are provided by an external waveform generator (Agilent 33220A). The equidistance was then evaluated as \cite{villares2014,delhaye2007}
\begin{equation}
 \epsilon=\frac{f_M-f_0}{M}-\frac{f_N-f_0}{N}.
 \label{eq:epsilon}
\end{equation}
Here $f_0$, $f_N$, and $f_M$ are the frequencies of three modes from the down-mixed multi-heterodyne spectrum. In the measurements reported the frequencies were given as $f_0=10$ MHz, $f_N=30$ MHz, and $f_M=70$ MHz with $N=1$, and $M=3$ as $\Delta f_{rep}=20$ MHz. In contrast to the experiments in ref. \cite{villares2014,delhaye2007} the combs are not stabilized. The fluctuations of the line at 70 MHz are shown in figure \ref{fig:HDBN}(b). These drifts have to be smaller than the bandwidth of the used filters (2 MHz). As the used configuration is very close to the noisy comb regime with beatnote linewidth of several hundred MHz, an active stabilization has proven to be very difficult to implement. Nevertheless, the equidistance of the comb modes can be verified. The histogram in figure \ref{fig:HDBN}(c) shows the deviation of $\epsilon$ (equation \eqref{eq:epsilon}). The values are fitted with a Gaussian curve with a center of -0.52 Hz and a standard deviation of 3.21 Hz. Normalized to the carrier frequency of 2.5 THz this corresponds to an accuracy of $1.3 \times 10^{-12}$. The Allan deviation of the experiment, normalized to the measurement bandwidth (3 times the free spectral range (FSR): 65.1 GHz), is shown in figure \ref{fig:HDBN}(d), showing a inverse square root dependency similar to Mid-IR QCL combs \cite{villares2014}, and Kerr combs \cite{delhaye2007}. \\
All these values could probably still be significantly improved by integrating longer and actively stabilizing one of the two combs. However our experiment shows that a verification of the equidistance of the comb modes down to the Hz-level is also possible in a free running mode of the combs. We are benefiting from the fact that the two combs share a big part of their noise as they are mounted on the same cryostat. \\
As a conclusion, we have reported a new approach for performing dual-comb experiments, with a measured frequency accuracy of $1.3 \times 10^{-12}$. Our approach fully exploits the unique characteristics of QCLs. The THz frequency signal is directly generated, and the detection does not require an external detector. Instead, we use one of the QCL combs of our dual-comb on-chip as a detector, realizing a fully integrated dual-comb system. One could also think of a future dual-comb setup with no need of recombining two combs on a detector but sending one comb through a sample (e.g. a gas cell) and then just focusing into a second comb acting as a detector. Such an approach would simplify the existing dual-comb setups to a high degree. The concept presented is not restricted to the THz but should also be implementable with Mid-IR QCL combs. \\
The experimental verification of the equidistance of the comb modes in THz QCL combs has been an important step in the development of compact and powerful THz combs. It proofs that QCL combs can be achieved at THz frequencies the same way they were first reported in the mid-IR \cite{Hugi2012}, i.e. with a broad and flat gain and no additional dispersion correction. Since the laser itself can be octave-spanning \cite{Roesch2014}, a dispersion correction on this laser could lead to an octave spanning dual-comb for high-accuracy THz spectroscopy down to a frequency accuracy of $10^{-12}$ or even below with an appropriate stabilization. Even though still at cryogenic temperatures such a dual-comb could pave a new way for spectroscopy at THz frequencies. 

\subsection*{Acknowledgement}
The presented work is part of the EU research project TERACOMB (Call identifier FP7-ICT-2011-C, Project No.296500). Additional funding comes from the Swiss National Science Foundation. The funding is gratefully acknowledged. The authors acknowledge the technical support of M.J. S\"uess and D. Kazakov.

\end{document}


\title{Supplementary Data: On-chip, self-detected THz dual-comb spectrometer}%

\author{Markus R\"osch}%
\email{mroesch@phys.ethz.ch}
\author{Giacomo Scalari}
\email{scalari@phys.ethz.ch}
\author{Gustavo Villares}
\author{Lorenzo Bosco}
\author{Mattias Beck}
\author{J\'er\^ome Faist}
\affiliation{ETH Zurich, Institute of Quantum Electronics, Auguste-Piccard-Hof 1, Zurich 8093, Switzerland}

\maketitle 

\begin{figure*}[t]
  \centering
  \includegraphics[scale=1]{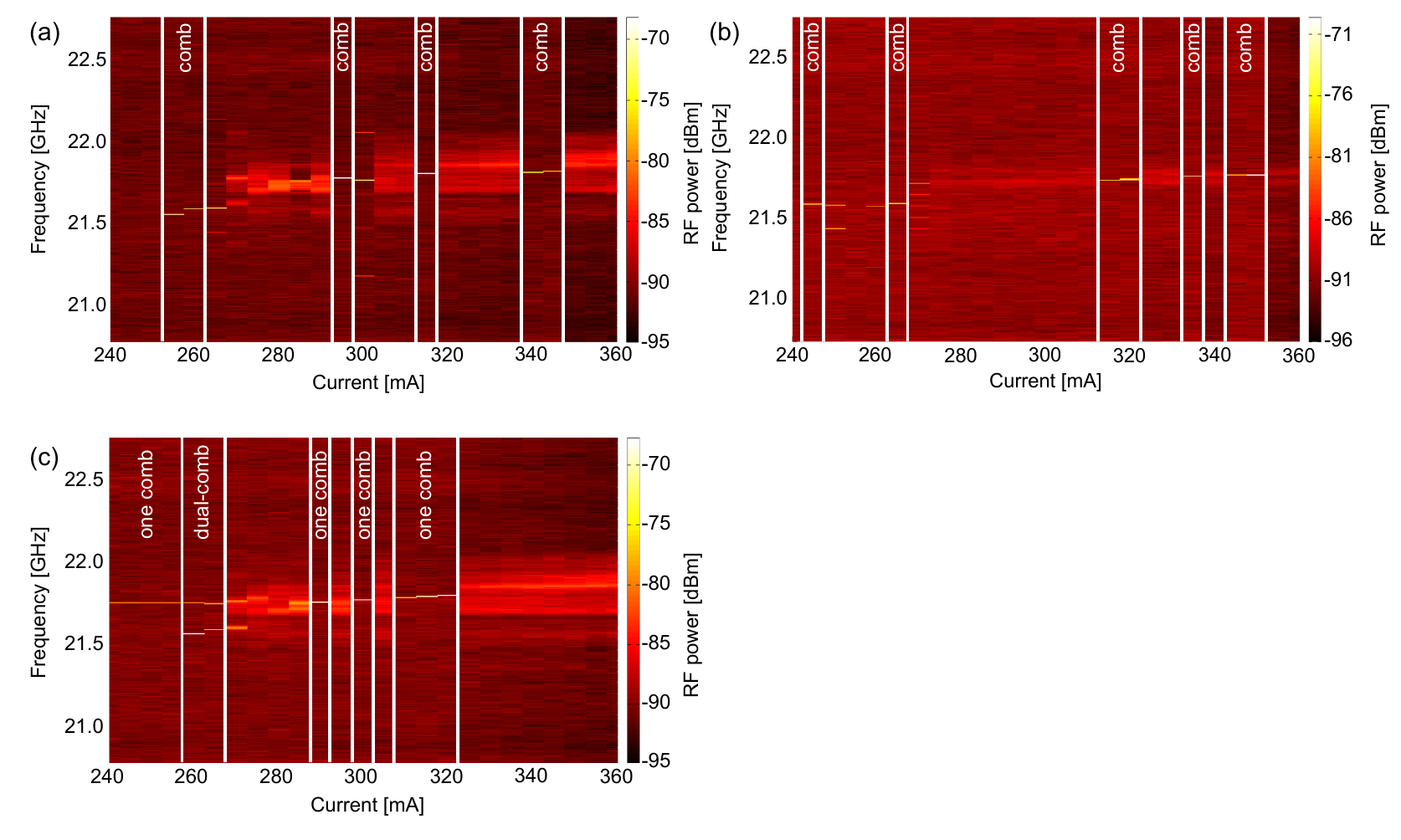}
  \caption{Beatnote characteristics: (a) intermode beatnote signal of laser 1 as function of driving current recorded over a span of 3 GHz with a RF spectrum analyzer. The laser temperature is fixed at 23 Kelvin. Laser 2 is switched off. (b) The same measurement but for laser 2 with laser 1 switched off. (c) The same scan as in (a) but with laser 2 fixed at 316mA, showing strong interaction of the two lasers. ((a)-(c): RBW=10 kHz, VBW=1 MHz, SWT=200 ms)}
  \label{fig:BNmap}
\end{figure*}

Both lasers have been characterized individually. Supplementary figure \ref{fig:BNmap}(a) and (b)  show a beatnote map for the two lasers. Similar to previous work with such lasers, the comb regime is limited to certain current ranges which are slightly different for different lasers \cite{Roesch2014,li2015}. The beatnote maps (supplementary figure \ref{fig:BNmap}(a), (b)) were measured in continuous wave (CW) at a fixed temperature of 23 Kelvin. Only the laser under test was switched on during the measurement. The laser was left free running without any active stabilization. A commercial power supply (Keithley instruments) has been used as a current source. As the current stability of this power supply is limited, a homemade low-pass filter has been employed to reduce fast current fluctuations on the laser bias. The beatnote was measured by picking up the RF signal from the current line. It is extracted with a bias-tee and the RF part is sent to a microwave spectrum analyzer (Rohde \& Schwarz FSU50). To acquire the beatnote maps (beatnote as a function of driving current) the spectrum analyzer was set with a resolution bandwidth (RBW) of 10 kHz and a video bandwidth (VBW) of 1MHz. The traces were recorded in a single sweep mode for every current step (5mA). The signal was amplified with a 35dB RF amplifier (Miteq AFS43-01002600-38-8P-44) and by another 35 dB using the internal preamplifier of the spectrum analyzer.\\
The operation conditions of the two lasers are not independent. This also affects the comb regimes of the lasers. When we are operating both lasers at the same time, the comb regimes eventually change. In supplementary figure \ref{fig:BNmap}(c) one laser was kept at a fixed current (316 mA) while the other laser was swept over the full current range. For low currents of the swept laser, the fixed laser is in a comb regime (Beatnote at 21.75 GHz in supplementary figure \ref{fig:BNmap}(c)). It leaves this regime when the swept laser is driven at a higher current ($>275$mA), although the temperature is fixed at a constant value of 23 Kelvin. But also the swept laser changes its behavior. For example the comb regime at 340mA is no longer there in the presence of the second laser.\\